# Ferrihydrite nanoparticles entrapped in shear-induced multilamellar vesicles

*Luigi Gentile[1,2*]*

1. Department of Chemistry, University of Bari "Aldo Moro", Via Orabona 4, Bari, 70126, Italy

2. Center of Colloid and Surface Science (CSGI) Bari Unit, Via Orabona 4, Bari, 70126, Italy

AUTHOR INFORMATION

**Corresponding Author**

*Luigi Gentile, Senior Lecturer, Department of Chemistry, University of Bari "Aldo Moro", 70125 Bari (Italy) orcid.org/0000-0001-6854-2963 email: luigi.gentile@uniba.it, Tel: +39 0805442033.

**Abstract**

*Hypothesis*

Ferrihydrite (Fh) nanoparticles are receiving considerable scientific interest due to their large reactive surface areas, crystalline structures, and nanoparticle morphology. They are of great importance in biogeochemical processes and have the ability to sequester hazardous and toxic substances. Here, the working hypothesis was to entrap fractal-like Fh nanoparticles, with a radius of gyration of 6.2 nm and a primary building block of polydisperse spheres with a radius of 0.8 nm, in a shear-induced multilamellar vesicle (MLV) state using a 40 wt.% polyethylene glycol dodecyl ether surfactant.

*Experiments*

Small- and Wide- Angle X-ray scattering revealed the equilibrium state of the non-ionic planar lamellar phase, the Fh dispersion, and their mixture. The MLV state was induced by using a shear flow in a Taylor-*Couette* geometry of a rheometer.

*Findings*

The nonionic surfactant initially exhibited a lamellar gel phase with two distinct *d*-spacings of 11.0 and 9.7 nm, which collapsed into the MLV state under shear flow. The Fh nanoparticles induced bilayer attraction by suppressing lamellar layer undulations, decreasing the *d*-spacing. These results are helpful in the understanding of the relationship between nanoparticle size and nanoparticle-bilayers interactions and provides insight on Fh encapsulations in a kinetically stable MLVs state.

## 1. Introduction

Ferrihydrite, Fh, is a hydrous ferric oxide mineral with a low degree of crystallinity copious in natural environments. Due to its large reactive surface area, it is involved in several biogeochemical processes.[1,2] Moreover, Fh is able to sequester and remove anionic dyes, which are toxic to living organisms.[3] Its crystalline structure has been debated due to its fine particle size and indeterminate structure and composition.[2,4] Recently, it has been found to consist of sheets of edge-sharing Fe-O octahedra, interspersed with sheets consisting of coexisting Fe-O octahedra and Fe-O tetrahedra, arranged in a hexagonal unit cell (space group P63mc).[5,6] X-ray diffraction has distinguished "two-line ferrihydrite" (two peaks in the patterns) as the least ordered form up to "six-line ferrihydrite" in its more ordered form. Here we focus on six-line ferrihydrite.

In nature and after synthesis, ferrihydrite occurs as nanoparticles ranging from 1 to 10 nm fully dispersed or aggregated in a fractal cluster.[6–8] The presence of surface groups in six-line Fh particles makes their shell water-rich, while their mineral core results hydrogen-poor.[6] pH and aging over time have a strong effect on Fh aggregates. A pH of 3.5 increases electrostatic repulsion, disaggregating clusters into their primary nanoparticles. On the other hand, aging leads to $CO_2$ adsorption, resulting in surface-adsorbed carbonate species[9] that prevent formation of metastable surface clusters toward the point of zero charge at around pH 8.[10] Nanoparticles are able to adsorb organic material at this interface[11] and their dynamics in polymer networks is the subject of intensive study.[12] Hiemstra[2] provided thermodynamic data for determining the solubility product for freshly prepared six-line Fh over a size range of 2.5-4 nm, $log([Fe^{3+}][OH^-]^3) = logK_{so} \sim 39.5$. The smallest particles in an Fh suspension react according to the Ostwald–Freundlich equation $RT\Delta lnK_{so} = 2/3\,\gamma A$, where A is the surface area and γ ($\approx 0.19$ J m$^{-2}$) is the interfacial tension [2], while the Fh suspension as whole reacts in line with the Ostwald equation $RT\Delta lnK_{so} = \gamma A$. The colloidal stability of Fh suspension is affected by several parameters such as initial size, pH, aging, and concentration that could lead to aggregation and eventually precipitation.

Here, an Fh nanoparticle dispersion was entrapped in a non-ionic lamellar phase to increase stability and to evaluate the effect on the lamellar phase. An aged six-line Fh dispersion (pH ~6.5) was used to prepare a 40 wt.% lamellar phase using tetraethylene glycol monododecyl ether ($C_{12}E_4$) with Mn ~362, known as Brij® L4 (L4). $C_nE_m$ surfactants are well known to form planar lamellar phase (Lα) morphology[13] even in mixed systems,[14] which under shear flow collapse into multilamellar vesicle (MLV) morphology; this is known as the Lα-to-MLV transition.[15–19] The *d*-spacing, i.e. layer spacing of the lamellae, for pure $C_{12}E_4$ at 40 wt.% in $D_2O$ is 6.8 nm.[20–22] However, Brij® L4 is not a pure $C_{12}E_4$ and may contain fractions of other $C_nE_m$.

## 2. Materials and methods

### 2.1 Materials

Materials. Brij® L4 nonionic surfactant made of polyethylene glycol dodecyl ether with Mn ~362 (i.e. ~ tetraethylene glycol monododecyl ether, $C_{12}E_4$), was purchased from Levanchimica Srl (Bari, Italy). The L4 lamellar phase was prepared by mixing 40 wt.% of the surfactant with distilled $H_2O$ (labeled L4-40wt). Six-line ferrihydrite (Fh) was synthesized as reported by Schwertmann and Cornell [1] by dissolving 18 g of $Fe(NO_3)_3 \cdot 9H_2O$ in 2 L of preheated distilled water by applying a continuous stirring. The suspension was kept at 75 °C for 11 min and then rapidly cooled to room temperature. The suspension was dialyzed against Milli-Q water to remove nitrates with a membrane cutoff of 12−14 kDa (Spectrum Laboratories Inc., Rancho Dominguez, USA) until the electrical conductivity was less than 5 μS m$^{-1}$. The solid concentration of Fh in the final nanoparticle dispersion was measure by solvent-evaporation and it was 1.6 g/l. The dispersion is affected by aging [7] and as such has been used after two years showing a final pH of ~6.5. The obtained dispersion was then diluted with $H_2O$ to obtain a 60 wt. % dispersion comparable with the nanoparticles concentration in the L4/ferrihydrite system (labeled L4Fh-40wt). L4-40wt and L4Fh-40wt systems were heated (~80°C) and stirred, cooling down to room temperature, to obtain a metastable lamellar phase.

### 2.2 Small- and Wide-Angle X-ray Scattering (SAXS and WAXS)

SAXS and WAXS measurements were performed using a pinhole-collimated system equipped with a Genix 3D X-ray source (Xenocs SA, Sassenage, France), namely SAXSLab Ganesha 300XL instrument (SAXSLAB ApS, Skovlunde, Denmark). The scattering intensity, *I(q)*, was recorded with the Pilatus detector (Dectris Ltd, Baden, Switzerland) located at three distinct distances from the sample, yielding a scattering vector range $0.0042$ Å$^{-1}$ $\leq q \leq 1.75$ Å$^{-1}$. Samples were loaded in a 1.5 mm diameter quartz capillary and then sealed (Hilgenberg GmbH, Malsfeld, Germany). An external JULABO thermostat (JULABO, Seelbach, Germany) fixed to 25 °C controlled the temperature. The two-dimensional (2D) scattering pattern was radially averaged using SAXSGui v2.15.01 software to obtain *I(q)*. The measured scattering curves were corrected for the background scattering.

### 2.3 Rheology

Step rate tests of 900 s were employed as transient experiments to induce a lamellar-to-MLV transition in the L4-40wt and L4Fh-40wt systems, respectively. Upward stepped ramps of shear rates in the range from 0.1 to 100 s$^{-1}$ identified the onset of the transition for both systems. Rheological measurements were carried out using the MCR302e stress-controlled rheometer (Anton Paar Gmbh, Graz, Austria) equipped with a Taylor-*Couette* geometry, *i.e.* concentric cylinder geometry, (inner diameter of 16.662 mm and a gap of 0.704 mm). The temperature was kept to 25 °C by a Peltier system.

## 3. Results and discussion

### 3.1 Structural features at the equilibrium

Figure 1 shows the joint small- and wide-angle X-ray scattering (SAXS and WAXS) profiles for a 40 wt.% L4 non-ionic lamellar phase (L4-40wt), the Fh dispersion diluted with 40 wt.% H$_2$O to achieve the same concentration as the mixed sample (Fh-60wt), and the 40 wt.% L4 non-ionic lamellar phase prepared in the Fh dispersion (L4Fh-40wt).

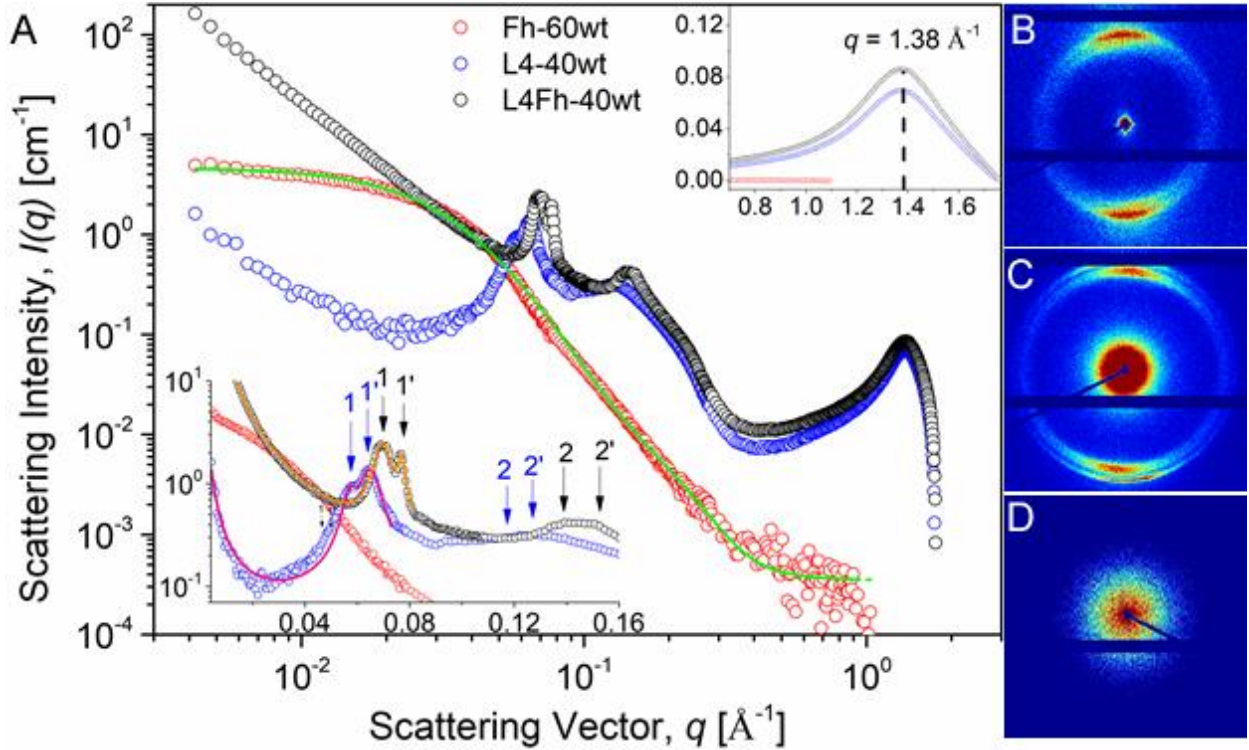

**Figure 1**. Radially averaged small- and wide-angle X-ray scattering (SAXS and WAXS) profiles (A) at 25 °C of 60 wt.% in water of the ferrihydrite nanoparticle dispersion (Fh-60wt; red dot), i.e. 0.96 g/l, 40 wt.% L4 nonionic surfactant in $H_2O$ (L4-40wt; blue dot), and 40 wt. % L4 nonionic surfactant in Fh nanoparticle dispersion (L4Fh-40wt; black dot). The green line shows the best fit for a fractal structure with a primary building block of polydisperse spheres, based on the Teixeira model,[23] eq. 1-3. The bottom-left inset shows a semi-log plot of the scattering profiles along with the best fit for the lamellar phase following Prevost et al.,[24] eq. 4, for the L4-40wt (pink) and L4Fh-40wt (orange) samples. The top-right inset shows a linear-linear plot of the WAXS profiles, with a peak at $q = 1.38$ Å$^{-1}$ for both the L4-40wt and L4Fh-40wt samples. 2D SAXS pattern in the low $q$-range for L4-40wt (B), L4Fh-40wt (C) and Fh-60wt (D).

The Fh dispersion scattering profiles are consistent with fractal-like aggregates with a primary building block of polydisperse spheres following the Teixeira model,[23] where the sphere radius is described by the Schulz distribution,

$$I(q)_{Fh} = P(q)_{pS} S(q)_{Teix} + bg \quad (1)$$

where $P(q)_{pS}$ is the form factor of the polydisperse sphere,

$$P(q)_{pS} = \phi V_p \Delta\rho^2 \int f_{Sch}(R_0, \bar{R}_0, \sigma) \left[\frac{3[sin(qR_0)] - qR_0 cos(qR_0)}{(qR_0)^3}\right]^2 dR_0 \quad (2)$$

and where $\phi$ is the volume fraction of the particles, $V_p$ is the volume of a single particle, $\Delta\rho$ is the scattering length density differences, $R_0$ is the particle radius, and $\sigma$ is the standard deviation of the Schulz distribution $f_{Sch}$, fixed here as 0.35. Based on the curve-fitting, $R_0$ = 0.8 nm, in agreement with Guénet et al.[25] and with the dynamic light scattering data in the supplementary material.

$S(q)_{Teix}$ is the Teixeira structure factor,[23]

$$S(q)_{Teix} = 1 + \frac{sin[(D-1)tan^{-1}(q\xi)]}{(qR_0)^D} \frac{D\Gamma(D-1)}{[1 + 1/(q^2\xi^2)]^{(D-1)/2}} \quad (3)$$

where $\Gamma$ is the gamma function, $D$ (equal to 3.2) is the fractal dimension, indicating a surface fractal, and $\xi$ is the correlation length of 2.4 nm. Thus, $R_g = \sqrt{D(D+1)\xi^2/2} = 6.2\ nm$, which is in line with the fractal aggregate size reported by Guénet et al.[25] of around 6 nm. However, these results differ from those of a previous study,[7] in which repulsive inter-particle interactions result in a structure factor which is different from 1 and a particle radius of 2.7 nm,[7] while yet other studies have reported a particle radius of 1.3 nm.[26–28] These differences may be due to higher adsorption of $CO_2$ on the nanoparticle surface when aged over a longer period (2 years) that results in a pH increase.

The L4 lamellar phase exhibited a peak in the WAXS region centered at $q_{WAXS}$ = 1.38 Å$^{-1}$, i.e. $d_{WAXS}$ = 0.455 nm, which implies lateral packing of alkyl chains perpendicular to the bilayer, indicating a gel lamellar phase $L_\beta$.[29–32] The $q_{WAXS}$ peak arises from the samples as shown in the supplementary material. Several models for fitting scattering data for lamellar phases have been suggested in the literature.[24,33–35] Here we adopted a model based on the Prevost *et al.*[24] approach, in which the scattered intensity arises from the lamellae stacking, described by a Lorentzian peak, with an initial increase due to diffuse small-angle scattering coming from local fluctuations in the surfactant concentration (layer undulation). However, two lamellar peaks can be observed in the 1D scattering profile (Figure 1) as well as in the 2D scattering profile, where two distinct pairs of Bragg peaks were detected, implying two lamellar phases. Moreover, two additional peaks were detected with a relative position ratio of 1:2 (bottom-left inset of Figure 1), due to the (100) and (110) reflections of the lamellar phase.[36,37] The presence of two lamellar spacing might be ascribed to the metastable state obtain during preparation. Further studies will be conducted on the equilibrium phase. The inter-lamellar distance (*d*-spacing) can be calculated from the *q* value of the first peak as $d_{L\beta 1} = 2\pi/q_{L\beta 1}^{peak}$ and $d_{L\beta 2} = 2\pi/q_{L\beta 2}^{peak}$, where $q^{peaks}$ is the center of the respective Bragg peaks. We have assumed that both lamellar phases, L$_{\beta 1}$ and L$_{\beta 2}$, are in the gel phase, although it is difficult to discriminate between them. Thus, the Prevost *et al.*[24] equation taking into account the two coexisting lamellar phases takes the following form,

$$I(q)_{L\alpha} = \frac{I_{L\beta 1}(0)}{1+(q\zeta_{L\beta 1})^2} + \frac{I_{L\beta 2}(0)}{1+(q\zeta_{L\beta 2})^2} + \frac{I_{L\beta 1}}{1+\left[\left(q-q_{L\beta 1}^{peak}\right)\zeta_{L\beta 1}^{peak}\right]^2}$$
$$+ \frac{I_{L\beta 2}}{1+\left[\left(q-q_{L\beta 2}^{peak}\right)\zeta_{L\beta 2}^{peak}\right]^2} \quad (4)$$

where $I_{L\beta1}(0)$, and $I_{L\beta2}(0)$ are intensity scaling factors and $I_{L\beta1}$ and $I_{L\beta2}$ are the Bragg peak intensities. $\zeta_{L\beta1}$ and $\zeta_{L\beta2}$ are the correlation lengths of the surfactant concentration fluctuations, while $\zeta_{L\beta1}^{peak}$ and $\zeta_{L\beta2}^{peak}$ are the inter-membrane correlation lengths (perpendicular to the membrane plane). As reported by Castro-Roman *et al.*[38] the layer undulations are taken into account in a non-self-consistent way in the Nallet model,[39] which is similar to the Prevost *et al.*[24] model. Here, the purpose is to determine the mean distance between the lamellar sheets. The L4 lamellar phase prepared in the Fh dispersion (L4Fh-40wt) exhibited Bragg peaks similar to the original L4 lamellar phase, while better defined at a higher *q*-position. The (110) reflections are also more pronounced than in the nanoparticles-free lamellar phase. At low-*q* a power law of 2.2 was observed, might due to smaller bilayer undulations than in the L4 prepared in water. However, the power law in the L4Fh-40wt sample may have been affected to some extent by fractal aggregation of the Fh nanoparticles. It is worth noting that no phase separation was observed several months after preparation. The two *d*-spacings obtained from eq. 4 for the L4 lamellar phase in water (L4-40wt) and L4 in Fh dispersion (L4Fh-40wt) are reported in Table 1.

**Table 1**. *d*-spacing $d_{L\beta1} = 2\pi/q_{L\beta1}^{peak}$ and $d_{L\beta2} = 2\pi/q_{L\beta2}^{peak}$ for the 40 wt.% L4 lamellar phase in water (L4-40wt) and in the ferrihydrite dispersion (L4Fh-40wt), where $q_{L\beta1}^{peak}$ and $q_{L\beta2}^{peak}$ were obtained from SAXS profile fitting (eq. 4), and where $I_{L\beta1}$ and $I_{L\beta2}$ are the relative intensities. The theoretical maximum inter-lamellar *d*-spacing was calculated using eq. 5. All values are in nm.

| | $d_{L\beta1}$ | $d_{L\beta2}$ | $I_{L\beta1}$ | $I_{L\beta2}$ | $d_{L\beta}^{max}$ |
|---|---|---|---|---|---|
| *L4-40wt* | 11.0 | 9.7 | 0.6 | 1.29 | 12.6 |
| *L4Fh-40wt* | 9.0 | 8.2 | 1.98 | 1.26 | 12.6 |

The theoretical maximum inter-lamellar *d*-spacing for an $L_\beta$ phase can be estimated as[40]

$$d_{max} = \frac{10^{24}\sqrt{3}}{d_{WAXS}{}^2 C_a N_a} \quad (5)$$

where $C_a$ is the total surfactant concentration in mol/l, $N_a$ is the Avogadro number, and $d_{WAXS}$ is the lattice constant obtained from the WAXS peak. The resulting $d_{max}$ is 12.6 nm. Considering that $I_{L\beta 1}$ and $I_{L\beta 2}$ are proportional to the relative amount of the two lamellar phases, the $L_\beta$ phase-swell was approximately 73% for the L4-40wt sample and 69% for the L4Fh-40wt sample. It is reasonable to assume that the Fh nanoparticles are located between the lamellar layers. Moreover, it has been demonstrated that nanoparticles can influence *d*-spacing in the lamellar phase, inducing bilayer attraction by suppressing layer undulations.[41–46] The power law of 2.2 at low-*q* for the L4Fh-40wt sample is a strong indication of this phenomenon.

**3.2 Entrapping Fh nanoparticles in the shear-induced MLVs**

The 40 wt.% L4 and L4Fh lamellar phases were subjected to shear flow using a Taylor-*Couette* geometry (gap = 0.704 mm) to induce the formation of multilamellar vesicles (MLVs) as reported for $C_{12}E_4$ at 40wt.%.[21,22] The steady-state viscosities obtained by transient experiments (Figure 2C and 2D) are shown in Figure 2A as a function of the shear rate, ranging from $10^{-1}$ to $10^2$ s$^{-1}$ for both the L4-40wt and L4Fh-40wt systems. The flow curve reveals three distinct regimes: the first regime represents slight shear thinning due to alignment of the lamellae in the flow direction[47], the second regime represents shear thickening due to MLV formation[15,17], and the third regime represents shear thinning due to the deformation of the densely packed MLV.[48] The maximum in the shear thickening region is shifted to a higher shear rate for L4Fh-40wt as can be seen also from the transient experiments (Figure 2D), indicating higher bending rigidity with a more pronounced viscosity maximum than that observed for the L4-40wt system. The onset of the Lα-to-MLV transition for the L4-40wt and the L4Fh-40wt is then 0.5 s$^{-1}$ and 5 s$^{-1}$, respectively. The corresponding shear rate-shear stress data (Figure 2B) reveal unusual behaviour for the L4Fh-40wt system that can be interpreted in terms of the Johnson–Segalman (JS) model similarly to Olmsted *et al.*[49,50] approach for rod particles, in which a possible phase separation *via* spinodal decomposition can lead to shear banding.[51–55] This phenomenon requires further investigation. Instead, the aqueous Fh nanoparticle dispersion shows Newtonian behavior (supplementary material).

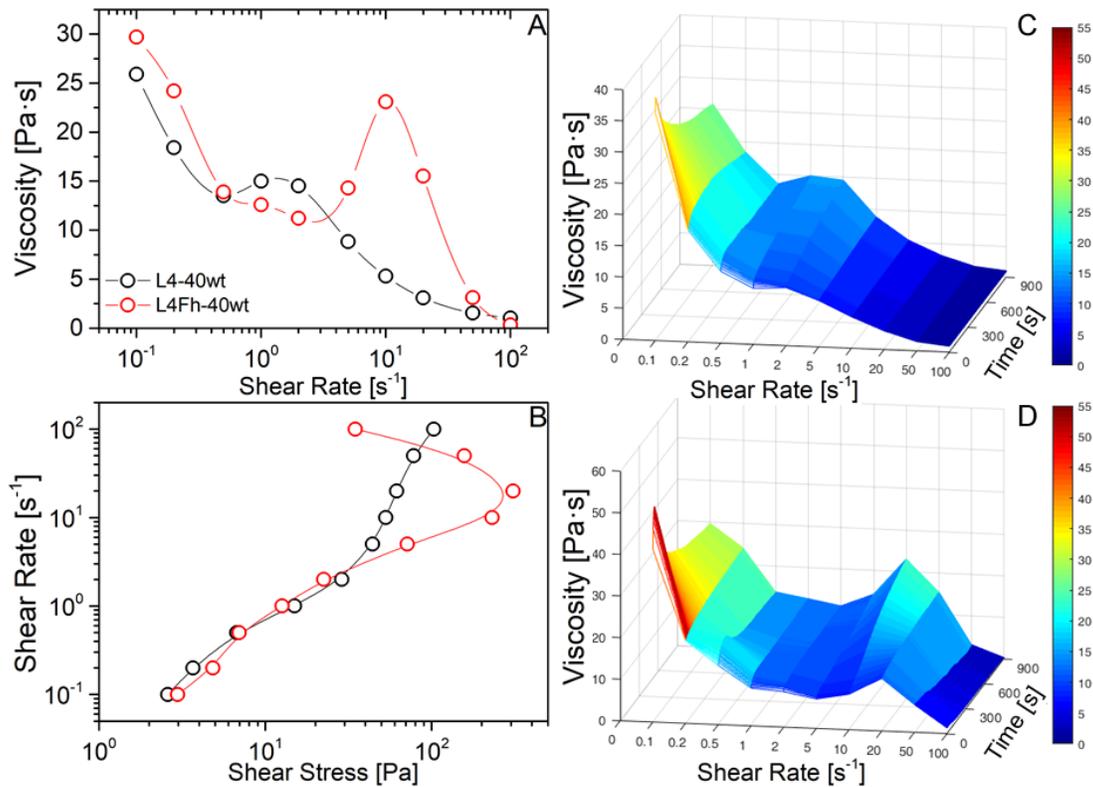

**Figure 2**. Steady-state viscosity as a function of the shear rate (A) and corresponding shear rate-shear stress plot (B) for the 40 wt.% L4 lamellar phase (L4-40wt) and the 40 wt.% L4 lamellar phase prepared in the Fh dispersion (L4Fh-40wt). Three-dimensional plots of the transient experiments performed to obtain the steady-state values are reported for both the L4-40wt (C) and L4Fh-40wt systems (D). The color scale indicates viscosity values from 0 (blue) to 55 (red) Pa s to facilitate comparisons between the two systems.

The frequency sweep of the planar lamellar phase before applying shear flow and the corresponding frequency sweep after applying a shear rate of 5 s$^{-1}$ are shown in Figure 3A and 3B, respectively, for the L4-40wt and L4Fh-40wt systems. The oscillatory measurement before the transient viscosity experiment (i.e. before applying shear flow) shows a storage modulus $G'$ higher than the loss modulus $G''$ over the frequency range of 1 to 100 rad/s. After the transient viscosity experiment at a shear rate of 5 s$^{-1}$, $G''$ increased only for the L4Fh-40wt system while $G'$ increased for both systems, 2-fold for the L4-40wt and 5-fold for the L4Fh-40wt systems. The larger gap between $G'$ and $G''$ and the corresponding shear thickening indicate the MLV formation.[56]

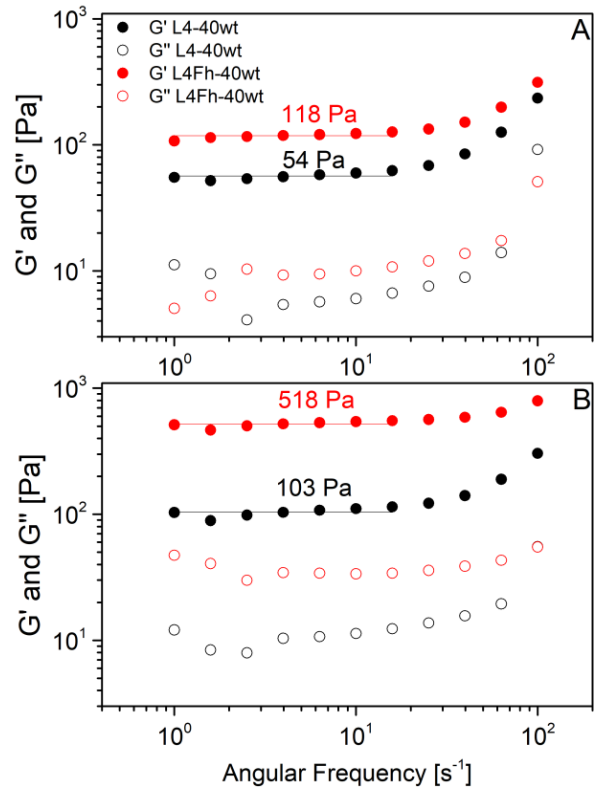

**Figure 3**. Frequency sweep measurements before (A) and after (B) applying a shear rate of 5 s$^{-1}$ to the 40 wt.% L4 lamellar phase (L4-40wt) and for the 40 wt.% L4 lamellar phase prepared in the Fh dispersion (L4Fh-40wt). The red and black lines are linear fitting of the data.

The key contribution to the elastic modulus $G'$ may arise from restricted motion due to vesicle-vesicle interaction along with the length scale corresponding to the MLV size or the $d$-spacing. To determine the effect of the $d$-spacing on the bilayer bending rigidity, $K$, we applied the approach of Colafemmina *et al.*[31],

$$K = \frac{4}{n} \frac{G'}{tan^2\theta} \bar{d}^3 \quad (6)$$

where here, $\bar{d}$ is the weighted average of the $d$-spacing with respect to the $I_{L\beta 1}$ and $I_{L\beta 2}$ scattering intensities, $2\theta$ is the top angle of the cone over which two neighboring MLVs interact ($tan^2\theta$ = 0.2), $n$ is the number of neighbours (~10), and $G'$ is the value obtained at low angular frequency from Figure 3. The bending rigidity of the L4-40wt in the MLV state is then ~0.05 k$_b$T, while the bending rigidity of the L4Fh-40wt is ~0.17 k$_b$T in the MLV state. However, the bending rigidity of the C$_{12}$E$_4$ planar lamellar phase is known to be ~4.2 k$_b$T.[57] The larger bending modulus of the planar lamellae than the MLV states is due to the loss of out-of-plane membrane fluctuations induced by the shear flow that opposes the Helfrich entropic repulsion, suggesting a loss of resistance to compression in the MLV states.[58] The mechanical tension of the membrane is then affected by the mean and spontaneous curvatures of the vesicle and by its shape.[59]

## 4. Conclusions

We demonstrate here that Fh nanoparticles can be entrapped in a Brij L4 metastable lamellar phase at 40 wt.%, generating a kinetically stable system.

The Fh nanoparticles affect the *d*-spacing of the lamellar phase, inducing bilayer attraction by suppressing layer undulations. This evidence is in agreement with a previous study of $C_{12}E_4$ and silica nanoparticles, demonstrating the presence of nanoparticles in the bilayer spacing.[45] These results are also consistent with simulations of nanoparticles entrapped in the lamellar phase of diblock copolymers,[60] which have proven that spherical nanoparticles increases the strength of the interaction between the nanoparticles and blocks. Moreover, increasing the nanoparticle radius and/or volume fraction affects the degree of order of the lamellar phase until a bicontinuous morphology is formed.[60] Other simulations on zwitterionic unilamellar vesicles have shown that nanoparticles with a low-density charge occupy the centre of the vesicle membrane.[61] However, when the surface charge density of the nanoparticles increases, the positively charged nanoparticles are repulsed by the vesicles.[61]

Here we highlighted for the first time that multilamellar vesicle formation under shear flow is affected by the presence of the nanoparticles in that a higher shear rate is needed to form the MLV state than for a lamellar phase free of nanoparticles. The bilayer bending rigidity of the lamellar phase in the MLV state is strongly affected by the presence of Fh nanoparticles. We have demonstrated that highly positively charged nanoparticles, but slightly larger than the *d*-spacing can be entrapped in a nonionic lamellar phase, indicating that their charge is compensated by the absorption of the alkyl carbon chain of the surfactant molecules in the bilayer thickness, an important insight in understanding the relationship between nanoparticle size and nanoparticle-bilayers interactions.

Based on these results, we envision that ferrihydrite nanoparticles could be delivered in a controlled way using the kinetic stability of the MLV state for a variety of purposes such as trapping toxic materials. Moreover, further studies on the probable shear banding phenomena and on the interplay with the non-ionic surfactants are of key interest.

ACKNOWLEDGMENT

I am grateful to Prof. Ulf Olsson (Lund University, Sweden) for discussions on small-angle scattering techniques and lyotropic liquid crystals and to Prof. Per Persson (Lund University, Sweden) for an introduction to iron-based nanoparticles. I thank "Fondazione Puglia" (Bari, Italy) for supporting the purchase of rheometer as well as Dr. Riccardo Di Ciuccio and Dr. Matera Sergio (Anton Paar, Torino, Italy) for the stimulating rheological discussions.

**Figure Captions**

**Figure 1**. Radially averaged small- and wide-angle X-ray scattering (SAXS and WAXS) profiles (A) at 25 °C of 60 wt.% in water of the ferrihydrite nanoparticle dispersion (Fh-60wt; red dot), i.e. 0.96 g/l, 40 wt.% L4 nonionic surfactant in $H_2O$ (L4-40wt; blue dot), and 40 wt. % L4 nonionic surfactant in Fh nanoparticle dispersion (L4Fh-40wt; black dot). The green line shows the best fit for a fractal structure with a primary building block of polydisperse spheres, based on the Teixeira model,[23] eq. 1-3. The bottom-left inset shows a semi-log plot of the scattering profiles along with the best fit for the lamellar phase following Prevost *et al.*,[24] eq. 4, for the L4-40wt (pink) and L4Fh-40wt (orange) samples. The top-right inset shows a linear-linear plot of the WAXS profiles, with a peak at $q = 1.38$ Å$^{-1}$ for both the L4-40wt and L4Fh-40wt samples. 2D SAXS pattern in the low $q$-range for L4-40wt (B), L4Fh-40wt (C) and Fh-60wt (D).

**Figure 2**. Steady-state viscosity as a function of the shear rate (A) and corresponding shear rate-shear stress plot (B) for the 40 wt.% L4 lamellar phase (L4-40wt) and the 40 wt.% L4 lamellar phase prepared in the Fh dispersion (L4Fh-40wt). Three-dimensional plots of the transient experiments performed to obtain the steady-state values are reported for both the L4-40wt (C) and L4Fh-40wt systems (D). The color scale indicates viscosity values from 0 (blue) to 55 (red) Pa s to facilitate comparisons between the two systems.

**Figure 3**. Frequency sweep measurements before (A) and after (B) applying a shear rate of 5 s$^{-1}$ to the 40 wt.% L4 lamellar phase (L4-40wt) and for the 40 wt.% L4 lamellar phase prepared in the Fh dispersion (L4Fh-40wt). The red and black lines are linear fitting of the data.

**Tables**

**Table 1**. $d$-spacing $d_{L\beta1} = 2\pi/q_{L\beta1}^{peak}$ and $d_{L\beta2} = 2\pi/q_{L\beta2}^{peak}$ for the 40 wt.% L4 lamellar phase in water (L4-40wt) and in the ferrihydrite dispersion (L4Fh-40wt), where $q_{L\beta1}^{peak}$ and $q_{L\beta2}^{peak}$ were obtained from SAXS profile fitting (eq. 4), and where $I_{L\beta1}$ and $I_{L\beta2}$ are the relative intensities. The theoretical maximum inter-lamellar $d$-spacing was calculated using eq. 5. All values are in nm.

|  | $d_{L\beta1}$ | $d_{L\beta2}$ | $I_{L\beta1}$ | $I_{L\beta2}$ | $d_{L\beta}^{max}$ |
|---|---|---|---|---|---|
| ***L4-40wt*** | 11.0 | 9.7 | 0.6 | 1.29 | 12.6 |
| ***L4Fh-40wt*** | 9.0 | 8.2 | 1.98 | 1.26 | 12.6 |

# Supplementary material

# Ferrihydrite nanoparticles entrapped in shear-induced multilamellar vesicles

*Luigi Gentile[1,2]\**

1. Department of Chemistry, University of Bari "Aldo Moro", Via Orabona 4, Bari, 70126, Italy

2. Center of Colloid and Surface Science (CSGI) Bari Unit, Via Orabona 4, Bari, 70126, Italy

AUTHOR INFORMATION

**Corresponding Author**

*Luigi Gentile, Senior lecturer at the Department of Chemistry of the University of Bari "Aldo Moro", 70125 Bari (Italy) orcid.org/0000-0001-6854-2963 email: luigi.gentile@uniba.it, Tel: +39 0805442033.

1. **Normalization of the wide-angle X-ray scattering data**

The 40 wt.% lamellar phase of a tetraethylene glycol monododecyl ether with Mn ~362 (L4-40wt) and the 40 wt.% lamellar phase prepared in the ferrihydrite dispersion (L4Fh-40wt) show a peak in the WAXS regime centered at $q_{WAXS}$ = 1.38 Å$^{-1}$, which implies a lateral packing of alkyl chain perpendicular to the bilayer indicating a gel lamellar phase $L_\beta$.[1–3] The peak is clearly coming from the lamellar phase as can be seen in Figure S1 where the H$_2$O (background) is compared with the ferrihydrite dispersion diluted to 60 wt.% (Fh-60wt), L4-40wt and L4Fh-40wt as such is not due to a wrong background subtraction.

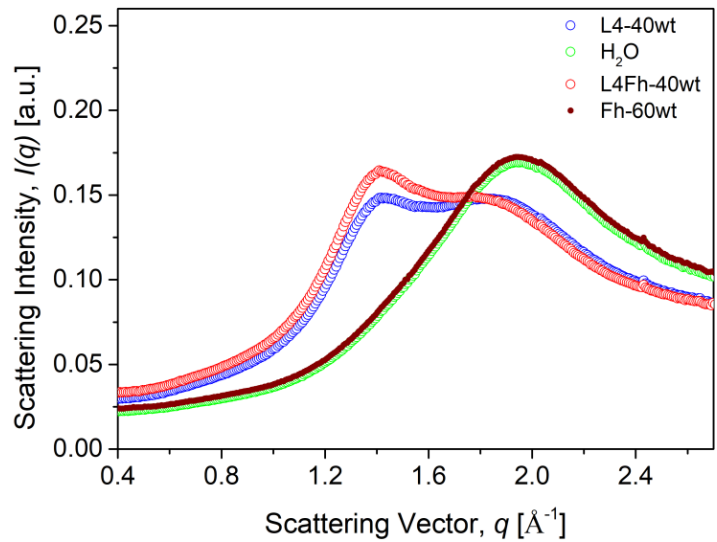

**Figure S1**. Wide-angle X-ray profiles in a linear-linear plot of the background (H$_2$O), ferrihydrite dispersion diluted in H$_2$O to 60 wt.% (Fh-60wt), 40 wt.% lamellar phase in H$_2$O (L4-40wt) and 40 wt.% lamellar phase prepared using the ferrihydrite dispersion (L4Fh-40wt).

2. **Dynamic light scattering and zeta-potential**

The Zetasizer Nano ZS instrument from Malvern Instruments, Ltd., Worcestershire, UK, was used for a dynamic light scattering (DLS) measurement at $\theta$ = 173° in addition to an electrophoretic mobility measurement of the Ferrihydrite dispersion. The main volume-weighted distribution was centered on the hydrodynamic radius, $R_H$, of 1.39 nm (Figure S2). The ratio between the radius of gyration, $R_g$, and $R_H$, for a solid sphere yields 0.775, i.e. $R_g \sim 1$ nm close to 0.8 nm of the spherical primary building block, $R_0$, of the fractal aggregate in the SAXS analysis (main text). The goniometer system was equipped with a 4-mW He-Ne laser with an automatic laser attenuator, and the detection unit comprised an avalanche photodiode. The temperature was set to 25 °C. The Ferrihydrite dispersion was filled in disposable folded capillary cells, and the measurements were performed at a fixed scattering angle of 173° using a laser interferometric technique (laser Doppler electrophoresis), which enabled the determination of the electrophoretic mobility. In such an experiment, an electric field is applied to a dispersion of charged particles that move with a velocity ($v = |\bar{v}|$), and the Doppler-shifted frequency of the incident laser beam caused by these moving particles is monitored. The velocity of a particle with radius R moving in an applied electric field, $E = |\bar{E}|$), is $v = u_e E$, where $u_e$ is the electrophoretic mobility. The zeta potential, ζ, was calculated from the Helmholtz-Smoluchowski equation.[4] The ζ was +55 mV.

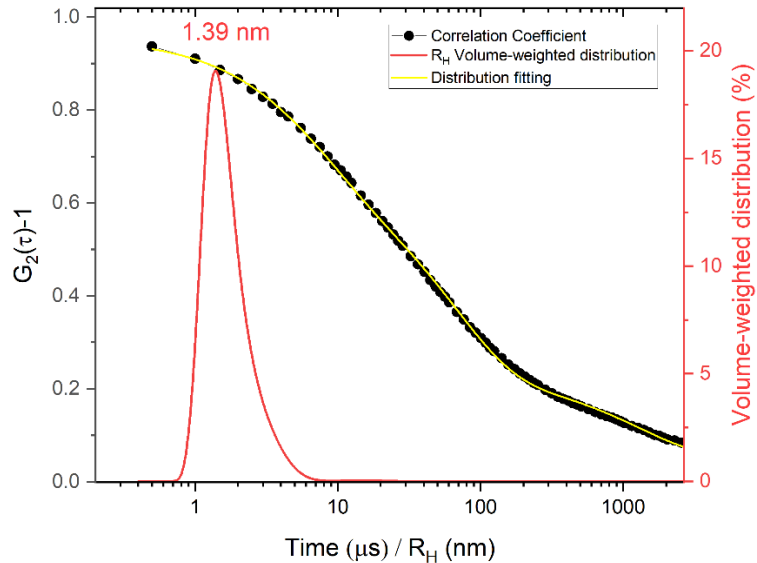

**Figure S2**. Correlation function and size distribution calculated by means of CONTIN algorithm for water Ferrohydrite nanoparticle dispersion.

### 3. Rheological behavior of Ferrihydrite dispersion

Figure S3 shows the Newotnian behaviour of the Ferrihydrite dispersion.

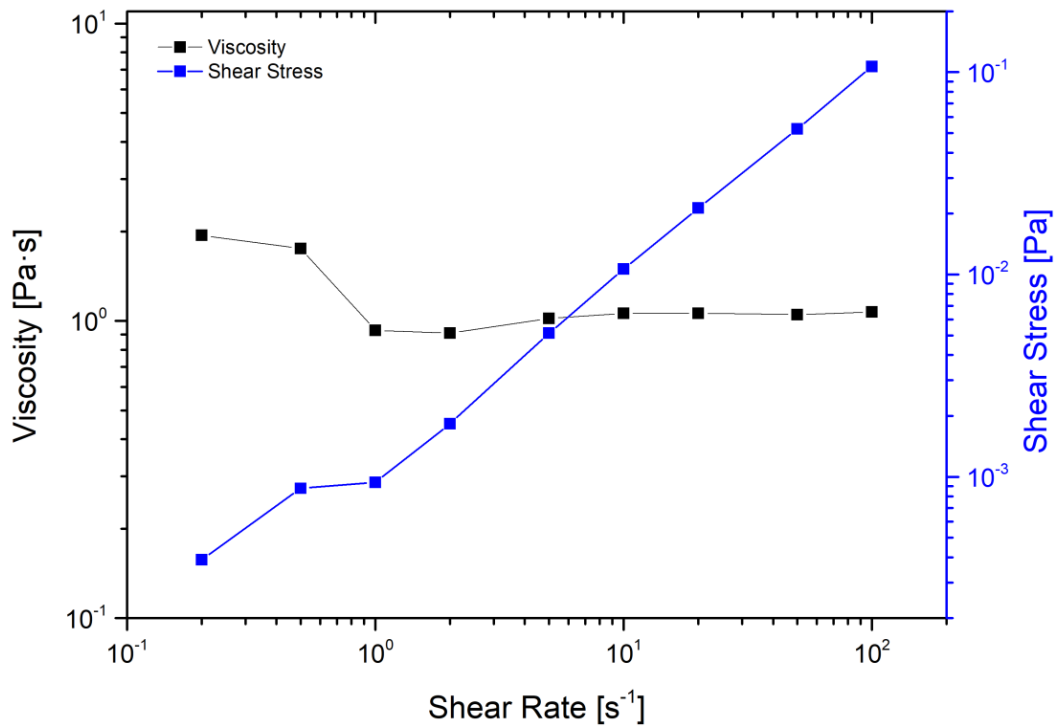

Figure S3. Steady-state viscosity and corresponding shear stress as a function of the shear rate for Ferrihydrate dispersion (Fh).